\documentclass[
  aps,
  prl,
  reprint,
  superscriptaddress,
  longbibliography,
  preprintnumbers
]{revtex4-2}

\usepackage[utf8]{inputenc}
\usepackage[T1]{fontenc}
\usepackage{amsmath,amssymb,mathtools,siunitx}
\usepackage{graphicx}
\usepackage{xcolor}
\usepackage{fontawesome}
\usepackage{xspace}
\usepackage{hyperref}
\usepackage[capitalize]{cleveref}
\usepackage[normalem]{ulem}

\hypersetup{
  colorlinks=true,
  linkcolor=blue!55!black,
  citecolor=blue!55!black,
  urlcolor=blue!55!black
}

\definecolor{DDcolor}{RGB}{8,81,156}       
\definecolor{DHcolor}{RGB}{49,130,189}     
\definecolor{DFcolor}{RGB}{107,174,214}    

\definecolor{IDcolor}{RGB}{165,15,21}      
\definecolor{IHcolor}{RGB}{222,45,38}      
\definecolor{IFcolor}{RGB}{251,106,74}     

\definecolor{MHmcolor}{RGB}{49,163,84}     
\definecolor{MHpcolor}{RGB}{117,107,177}   

\newcommand{\DDcol}[1]{\textcolor{DDcolor}{#1}}
\newcommand{\DHcol}[1]{\textcolor{DHcolor}{#1}}
\newcommand{\DFcol}[1]{\textcolor{DFcolor}{#1}}

\newcommand{\IDcol}[1]{\textcolor{IDcolor}{#1}}
\newcommand{\IHcol}[1]{\textcolor{IHcolor}{#1}}
\newcommand{\IFcol}[1]{\textcolor{IFcolor}{#1}}

\newcommand{\MHmcol}[1]{\textcolor{MHmcolor}{#1}}
\newcommand{\MHpcol}[1]{\textcolor{MHpcolor}{#1}}

\graphicspath{{plots/}}

\newcommand*{\dd}{\mathrm{d}}
\newcommand*{\myendmatterref}{\hyperref[end-matter]{End Matter}}

\newcommand*{\addemdash}[1]{#1---}
\makeatletter
\renewcommand{\paragraph}{%
    \@startsection%
    {paragraph}%
    {4}%
    {\parindent}%
    {\z@}%
    {0pt}%
    {\normalfont\normalsize\itshape\addemdash}%
}
\makeatother

\begin{document}


\preprint{DESY-26-162}
\preprint{IFT-UAM/CSIC-26-166}
\preprint{IPPP/26/84}

\title{Microscopic Realizability of Non-unique Bubble Hydrodynamics}

\author{Giulio Barni}
\email{giulio.barni@desy.de}
\affiliation{Instituto de F\'isica Te\'orica IFT-UAM/CSIC, Cantoblanco, E-28049 Madrid, Spain}
\affiliation{Deutsches Elektronen-Synchrotron DESY, Notkestr.~85, 22607 Hamburg, Germany}

\author{Eric Madge}
\email{eric.madge-pimentel@durham.ac.uk}
\affiliation{Instituto de F\'isica Te\'orica IFT-UAM/CSIC, Cantoblanco, E-28049 Madrid, Spain}
\affiliation{Departamento de F\'isica Te\'orica, Universidad Aut\'onoma de Madrid, Cantoblanco, E-28049 Madrid, Spain}
\affiliation{Institute for Particle Physics Phenomenology, Durham University, Durham DH1\,3LE, United Kingdom}

\begin{abstract}
    First-order phase transitions drive fluid motion that induces energy transport and gravitational-wave production. 
    The flows are usually classified in terms of the wall velocity and the transition strength.
    We show that this description fails when the two phases have unequal sound speeds $c_\pm$.
    New mixed-hybrid profiles may appear, and for a strong hierarchy, $c_+\gg c_-$, where the phase behind the wall is much softer, heating or cooling ahead of the wall allow for the coexistence of distinct stationary flows at the same wall velocity and transition strength.
    This ambiguity is removed once thermodynamic normalization and entropy production are specified.
    We identify microscopic settings that realize these sound-speed hierarchies.
    In thermal plasmas it requires exponentially large mass-to-temperature ratios, whereas at finite density a conserved charge removes this obstruction, making dense matter its natural arena.
\end{abstract}

\maketitle

\paragraph{Introduction}

First-order phase transitions proceed through the nucleation and collision of bubbles of one phase within the other, linking microscopic properties of matter to macroscopic fluid dynamics~\cite{Langer:1969bc,Landau:1980mil,Bauswein:2018bma,Hindmarsh:2020hop}.
As the bubbles expand, their walls drive fluid motion and redistribute energy between heat and bulk flow~\cite{Langer:1969bc,Coleman:1977py,Linde:1980tt,Linde:1981zj}.

When microscopic length scales are small relative to the bubble radius, the wall can be treated as a thin interface separating locally equilibrated fluids.
For a wall moving at constant velocity, the spherically symmetric flow approaches a self-similar profile at late times, with shocks and rarefaction waves determined by energy--momentum conservation~\cite{Steinhardt:1981ct,Ignatius:1993qn,Laine:1993ey,Kurki-Suonio:1995rrv,Espinosa:2010hh}.
In cosmological transitions, these profiles seed acoustic motion and gravitational-wave production~\cite{Hindmarsh:2013xza,Hindmarsh:2015qta,Hindmarsh:2017gnf,Jinno:2020eqg,Jinno:2022mie}.
The sound-shell model provides a semi-analytic link between the profiles and the resulting spectrum~\cite{Hindmarsh:2016lnk,Hindmarsh:2019phv,Guo:2020grp,Cai:2023guc,RoperPol:2023dzg,Giombi:2024kju,Giombi:2025tkv,Barni:2025gnm,Caprini:2026nnk}.

In the standard treatment, each phase is described by the bag equation of state, which models it as radiation plus a constant zero-temperature vacuum energy.
Both phases then have the conformal sound speed $c_s=1/\sqrt{3}$, and stationary profiles are uniquely determined by the wall velocity $\xi_w$ and the transition strength $\alpha_N$ in the unperturbed fluid~\cite{Espinosa:2010hh,Barni:2024lkj}.
However, mass thresholds and strong interactions can drive the equation of state away from conformality, motivating generalizations with independent sound speeds $c_+$ and $c_-$ that alter the allowed profiles and their energy budget~\cite{Leitao:2014pda,Giese:2020rtr,Giese:2020znk,Tenkanen:2022tly,Wang:2022lyd,Wang:2023jto,Tian:2024ysd,Mateos:2026moc}.

Stationary profiles are usually divided into direct solutions, in which the fluid, as seen from the bubble rest frame, moves along with the wall~\cite{Espinosa:2010hh}, and inverse solutions, in which it moves against it~\cite{Barni:2024lkj}.
As we discuss in detail in a companion paper~\cite{Barni:2026hydrodex}, for $c_+>c_-$ the classification admits a third class, which we dub mixed hybrids, in which the fluid flow reverses direction across the wall.
For a sufficiently large hierarchy, $c_+\gg c_-$, distinct self-similar profiles can coexist at the same $(\xi_w,\alpha_N)$, making the classification in terms of these inputs non-unique.
At still larger hierarchies, inverse solutions can extend into the $\alpha_N>0$ region, populated only by direct solutions with outward flow in the bag model.

Resolving this ambiguity requires additional information, such as the relative thermodynamic normalization of the phases, which encodes their effective thermal degrees of freedom, together with a prescription for entropy production at the wall~\cite{Ai:2021kak,Ai:2023see,Sanchez-Garitaonandia:2023zqz,Krajewski:2024zxg}. 
In this Letter, we further assess whether microscopic systems can indeed realize the sound-speed hierarchies for which the classification is ambiguous.
We find that it is difficult to achieve in thermal and confining plasmas, whereas a conserved charge generates it naturally, identifying dense matter as the most promising setting.

\paragraph{Fluid profile solutions}

\begin{figure*}[t]
  \centering
  \includegraphics[width=0.97\textwidth]{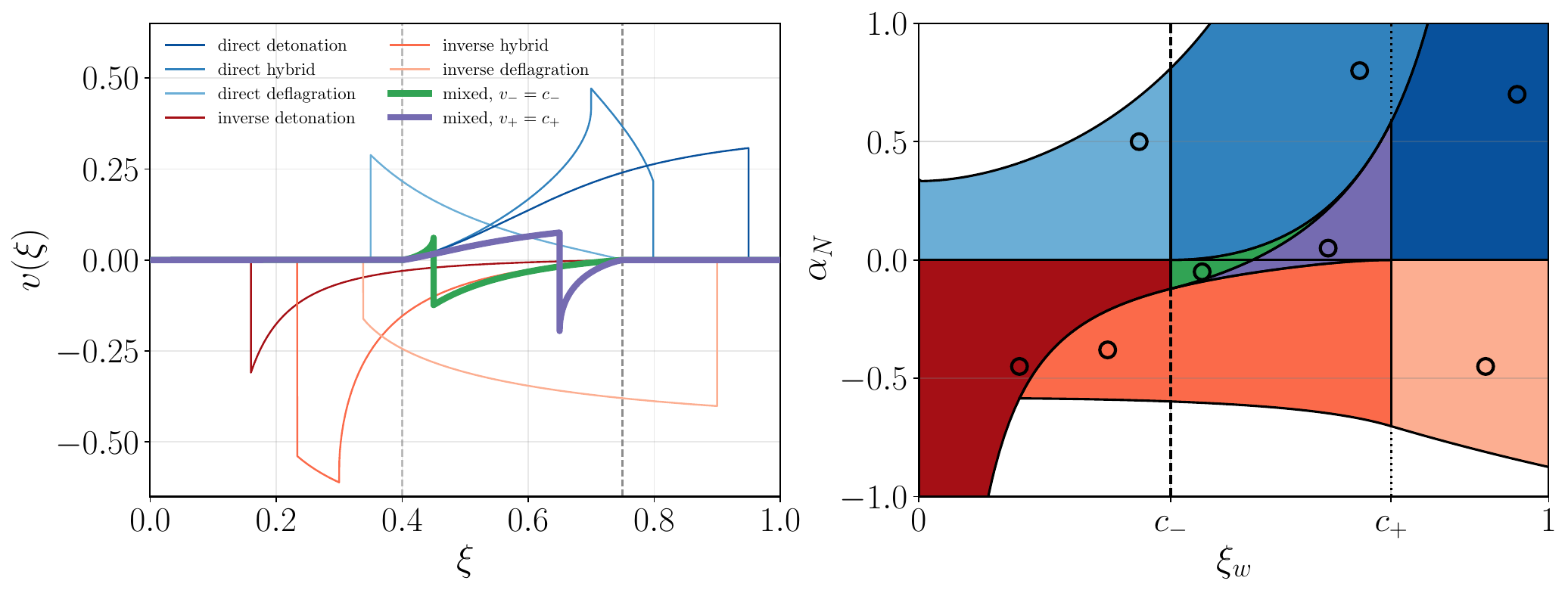}
  \caption{%
    \textbf{Hydrodynamic profiles beyond the bag model.}
    Left:
    representative fluid velocity profiles $v(\xi)$ in the bubble-center frame for the different stationary branches:
    \DDcol{direct detonations}, \DHcol{direct hybrids}, \DFcol{direct deflagrations}, \IDcol{inverse detonations}, \IHcol{inverse hybrids}, \IFcol{inverse deflagrations}, \MHmcol{mixed-hybrids with $v_-=c_-$}, and \MHpcol{mixed hybrids with $v_+=c_+$}.
    Right:
    the corresponding domains in the $(\xi_w,\alpha_N)$ plane for $c_+=0.75$ and $c_-=0.40$, using the same color coding.
    Circles mark the parameters of the profiles displayed in the left panel.
    }
  \label{fig:classification}
\end{figure*}

We describe each phase by a constant-sound-speed (CSS) equation of state~\cite{Leitao:2014pda,Giese:2020rtr,Giese:2020znk},
\begin{align}
 e_\pm(T)&=a_\pm T^{q_\pm}+\epsilon_\pm, &
 p_\pm(T)&=c_\pm^2 a_\pm T^{q_\pm}-\epsilon_\pm,
\label{eq:css}
\end{align}
where $q_\pm=1+c_\pm^{-2}$.
Here, `$+$' and `$-$' denote the phases ahead of and behind the wall, respectively, $a_\pm$ fix their thermal normalizations and $\epsilon_\pm$ their zero-temperature vacuum energies.
The bag model follows for $c_+=c_-=1/\sqrt{3}$.
Denoting the fluid velocities immediately in front of or behind the wall in its rest frame by $v_\pm$, energy--momentum conservation across the interface gives
\cite{Steinhardt:1981ct,Espinosa:2010hh}
\begin{equation}
 v_+ v_-=\frac{p_+ - p_-}{e_+ - e_-},
 \qquad
 \frac{v_+}{v_-}=\frac{e_- + p_+}{e_+ + p_-}.
 \label{eq:matching}
\end{equation}
Boosting into the bubble-center frame, and focusing on the late time dynamics, the flow is self-similar, depending only on $\xi\equiv r/t$, with $r$ the distance from the nucleation point and $t$ the time since nucleation, and obeys 
\begin{align}
  \frac{\dd v}{\dd\xi} = \frac{2 \, c_s^2 v (1 - v^2) (1- \xi v)}{\xi [(\xi-v)^2 - c_s^2 (1-\xi v)^2]} \,,
  \label{eq:diff-eq}
\end{align}
where $c_s=c_\pm$ in front of and behind the wall, respectively.

Profiles with a rarefaction wave on one side and unperturbed fluid on the other are detonations;
replacing the rarefaction by a compression wave ending at a shock gives deflagrations.
Ordinary hybrids combine a shock and a rarefaction on opposite sides of the wall~\cite{Laine:1993ey,Kurki-Suonio:1995rrv,Espinosa:2010hh,Barni:2024lkj}, with their direct and inverse branches satisfying $v_-=c_-$ and $v_+=c_+$, respectively.
We classify solutions as direct or inverse according to whether the fluid flows along or opposite to the wall’s direction of propagation.

As pointed out in Ref.~\cite{Barni:2026hydrodex}, for $c_-<c_+$, a new profile class arises.
The interval $c_-<\xi_w<c_+$ supports solutions with rarefactions on both sides of the wall, with the inner fluid moving outward and the outer fluid inward.
These mixed hybrids have two branches, distinguished by which side satisfies the condition $v_\pm=c_\pm$.
\Cref{fig:classification} displays representative profiles and their domains.
A detailed classification is given in Ref.~\cite{Barni:2026hydrodex}.

To express these solutions in terms of the unperturbed fluid, we use the pseudo-trace strength~\cite{Giese:2020rtr,Giese:2020znk,Ai:2023see},
\begin{equation}
 \alpha_{\bar\theta}(T)=\frac{\Delta \bar\theta(T)}{3w_+(T)},
 \qquad \bar\theta=e-\frac{p}{c_-^2},
 \label{eq:strength}
\end{equation}
where $w=e+p$ is the enthalpy density and $\Delta \bar\theta(T)=\bar\theta_+(T)-\bar\theta_-(T)$ compares the phases at the same temperature.
The matching conditions in \cref{eq:matching} involve $\alpha_+\equiv\alpha_{\bar\theta}(T_+)$, whereas the unperturbed-fluid input is $\alpha_N\equiv\alpha_{\bar\theta}(T_N)$.
These can differ because the fluid ahead of the wall is heated or cooled.
Substituting \cref{eq:css} gives
\begin{align}
    \alpha_N&= r_w\, \alpha_+ + (1-r_w)\, \frac{c_-^2-c_+^2}{3c_-^2(1+c_+^2)},
\label{eq:projection}
\end{align}
where $r_w \equiv w_+(T_+)/w_+(T_N)$ measures the change in enthalpy caused by the fluid response in front of the wall, with $r_w>1$ for heating and $r_w<1$ for cooling.
An unperturbed upstream fluid has $r_w=1$ and $\alpha_+=\alpha_N$.

\paragraph{Degeneracies in the parametrization}

\begin{figure*}[t]
  \centering
  \includegraphics[width=0.96\textwidth]{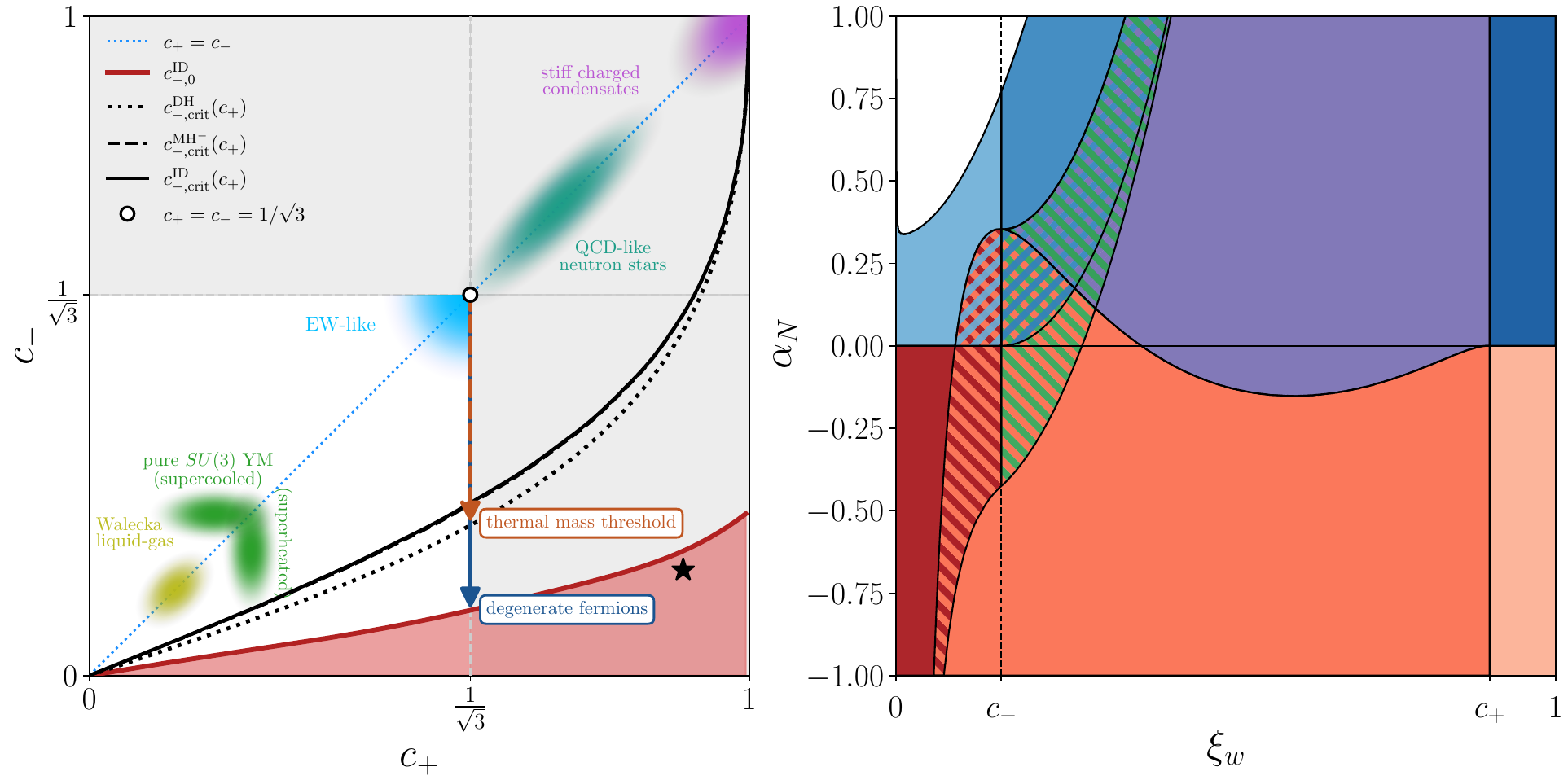}
  \caption{%
    \textbf{Left}:
    critical curves for the onset of folds on the ID (solid), MH$^-$ (dashed), and DH (dotted) branches in the $(c_+,c_-)$ plane.
    Folds occur below the corresponding curves.
    The red region marks the appearance of inverse solutions with $\alpha_N>0$.  
    Shaded patches and arrows indicate the schematic sound-speed ranges and parametric trends of the microscopic examples discussed in the text.
    The star marks $(c_+,c_-)=(0.9,0.16)$.
    \textbf{Right}:
    hydrodynamic domains in the $(\xi_w,\alpha_N)$ plane for this benchmark, with colors as in \cref{fig:classification} and hatching denoting regions where several solutions coexist.
  }
\label{fig:fold}
\end{figure*}

A unique profile at fixed $(\xi_w,\alpha_N)$ requires that distinct wall states remain distinguishable after accounting for the fluid disturbance ahead of the wall.
This does not need to hold as different local strengths can be compensated by different amounts of heating or cooling, yielding the same strength in the unperturbed fluid.

Along a given branch, at fixed $\xi_w$ and $c_\pm$,
degeneracies occur when the derivative $\partial_{v_+}\alpha_N|_{\xi_w}$ changes sign.
The map $(v_+,v_-) \mapsto (\xi_w, \alpha_N)$ folds, i.e.\ it becomes non-injective and turns back, and multiple wall states acquire the same $\alpha_N$.
As folds originate at the lower $\alpha_N$ boundary of a branch, they produce both intra-branch degeneracies and overlaps with other branches.

A complete branch-by-branch analysis in Ref.~\cite{Barni:2026hydrodex} proves that folds can occur only for $c_+>c_-$ and only on the direct hybrid (DH), inverse detonation (ID), and $v_-=c_-$ mixed-hybrid (MH$^-$) branches.
The threshold value of $c_-$ below which the DH branch develops a fold can be derived analytically~\cite{Barni:2026hydrodex},
\begin{equation}
 c_{-,\mathrm{crit}}^{\mathrm{DH}}(c_+)
 =\frac{3-c_+^2-\sqrt{(1-c_+^2)(9-c_+^2)}}{2c_+},
 \label{eq:dhcrit}
\end{equation}
while the corresponding thresholds on the MH and ID branches need to be determined numerically, satisfying
$c_{-,\mathrm{crit}}^{\mathrm{DH}}<c_{-,\mathrm{crit}}^{\mathrm{MH}^-}<c_{-,\mathrm{crit}}^{\mathrm{ID}} < 1/c_+-\sqrt{1/c_+^2-1}$.

When folds start to form, degeneracies are first limited to narrow intervals close to the borders between solution classes.
As the sound-speed hierarchy further increases, the regions of overlap grow.
As an indicator for when overlaps become significant, we take the threshold value $c_{-,0}^{\mathrm{ID}} (< c_{-,\mathrm{crit}}^{\mathrm{DH}})$ below which the ID branch (as well as inverse hybrids) crosses into the $\alpha_N>0$ region, which for $c_- \geq c_+$ is only populated by direct solutions.

The sound-speed hierarchies which exhibit degeneracies are depicted in \cref{fig:fold}.
The left panel maps the onset and subsequent development of the folds in the $(c_+,c_-)$ plane.
The critical curves in black mark where the ID, MH\textsuperscript{$-$}, and DH branches first become non-monotonic, with folds occurring below the lines.
Deeper into the hierarchical region, inverse solutions extend to $\alpha_N>0$, shaded in red.
The right panel shows the resulting classification for the
benchmark $(c_+,c_-)=(0.9,0.16)$, marked by the star.
Hatched regions, concentrated around $\xi_w\simeq c_-$, are covered by more than one branch.
IDs overlap among themselves and with DHs, while MHs\textsuperscript{$-$} overlap with both MHs\textsuperscript{$+$} and IHs.
Up to three profiles coexist at fixed $(\xi_w,\alpha_N)$, and inverse solutions reach $\alpha_N\simeq0.35>0$, deep in the region populated only by direct solutions in the bag model.

The degeneracies arise because $(\xi_w,\alpha_N,c_+,c_-)$ specifies the hydrodynamic solution only incompletely.
The missing thermodynamic input is $\Psi_N\equiv w_-(T_N)/w_+(T_N)$, which fixes the relative normalization of the phases and encodes their effective thermal degrees of freedom~\cite{Ai:2023see,Sanchez-Garitaonandia:2023zqz}.
Each candidate profile requires a definite value $\Psi_{N,{\rm LTE}}$ when entropy is conserved across the wall in local thermal equilibrium (LTE)~\cite{Ai:2021kak,Ai:2023see}.
For the benchmark above, these values are numerically distinct throughout the interiors of the coexistence regions~\cite{Barni:2026hydrodex}.
Fixing $\Psi_N$ therefore selects at most one LTE solution. 
The same reasoning extends beyond LTE, with the entropy flux changing by a prescribed ratio $\mathcal{R}$ with respect to the LTE value across the wall~\cite{Krajewski:2024zxg,Krajewski:2026kcm} (cf.\ \myendmatterref).

Without this additional information, heating, kinetic energy and the acoustic source are branch dependent. 
Nevertheless, sizeable overlaps require a pronounced sound-speed hierarchy, difficult to obtain in conventional radiation-dominated thermal plasmas~\cite{Giese:2020znk,Tenkanen:2022tly}.
The practical question is therefore which microscopic systems can reach this regime.

\paragraph{Can the required hierarchy be realized?}

Folds require the new phase, behind the wall, to be much softer than the old one, for supercooled and superheated transitions alike. 
For $c_+=1/\sqrt3$ folds appear at $c_+/c_-\simeq\numrange{2}{2.5}$ and extended overlaps at $\simeq\numrange{5}{6}$.
For larger $c_+=0.9$, already $c_+/c_-\simeq1.7$ suffices.
Near-conformal, electroweak-like plasmas (cyan in \cref{fig:fold}) remain far from this regime.
Realizing the hierarchy requires the new phase to be soft, and it must still carry the entire enthalpy flux entering the wall, $w_-\gamma_-^2v_-=w_+\gamma_+^2v_+$.
As we show next, these conditions compete in thermal plasmas but not at finite density.

Now, consider a real scalar $\phi$ with a tree-level barrier between minima at $\phi=0$ and $\phi=v$, coupled to $N_{\!f}$ Dirac fermions $\psi$ through $-y\phi\bar\psi\psi$.
For $m_\phi\gg T$ the scalar quanta are not thermally populated and $\phi$ only forms the wall.
The plasma thus consists of fermions, massless ahead of the wall and with mass $m=yv$ behind it.
The critical temperature follows from pressure equality, with the vacuum-energy difference compensating the thermal pressure lost by the fermions as they become heavy, $\epsilon_+-\epsilon_-\simeq4N_{\!f}T_c^4/\pi^2$.
For a Boltzmann gas,  $w_+=16 N_{\!f} T_+^4/\pi^2$ and $c_+^2=1/3$ ahead of the wall, while behind the wall~\cite{Landau:1980mil,Kapusta:2006pm},
\begin{equation}
 w_-=\frac{4\,N_{\!f}\,m^4}{2\pi^2}\frac{K_3(x)}{x},\qquad
 c_-^2=\frac{K_3(x)}{xK_2(x)+3K_3(x)},
\label{eq:nr-cs}
\end{equation}
where $x\equiv m/T_-$ and $K_\nu$ are modified Bessel functions of the second kind.
For $x\gg 1$, $c_-^2\to1/(x+3)$.
The flux condition fixes $w_-/w_+\sim\mathcal O(1)$ and hence $m/T_-\simeq4\ln(m/T_+)$, independently of $N_f$, so that
\begin{equation}
 \frac{c_+}{c_-}\simeq\left[\frac43\ln\frac{m}{T_+}\right]^{1/2}.
\label{eq:thermal-bound}
\end{equation}
The heavy gas must heat up until it carries the incoming flux, so the hierarchy grows only as the square root of a logarithm.
The fold onset requires $m/T_+\simeq70$, and $c_+/c_-\simeq5$ requires $m/T_+\sim10^9$ and $T_-/T_+\sim10^7$, with the broken phase superheated behind the wall and kept metastable only by the tree-level barrier.
Light species make things worse.
For a mixture of species $i$ with entropy densities $s_i$ and individual sound speeds $c_{s,i}$, the total sound speed obeys $c_s^{-2}=\sum_i(s_i/s)\,c_{s,i}^{-2}$.
The phase is therefore soft only if the nonrelativistic species, which have $c_{s,i}^2\simeq T/m_i\ll1/3$, dominate the entropy.
Already $c_+/c_-=2$ then requires $\gtrsim10^4$ heavy degrees of freedom per light one.

Confinement, a nonperturbative source of softness, faces a similar limit.
In $SU(3)$ Yang--Mills theory, lattice data give $c_s^2\simeq\numrange{0.01}{0.07}$ for the deconfined plasma within $2\%$ of $T_c$ and $c_s^2\simeq\numrange{0.05}{0.08}$ for the confined phase~%
\footnote{
    We compute $c_s^2=s/(T\,ds/dT)$ from the lattice entropy density $s(T)$ of Ref.~\cite{Giusti:2025fxu}, which is given in both phases on the side of $T_c$ where each is stable.
    We take the derivative of their fitted $s(T)$ and check it against the derivative estimated directly from neighbouring lattice points.
    For the deconfined phase this reproduces the $c_s$ shown in Ref.~\cite{Giusti:2025fxu}.
    No lattice data exist for the metastable phases.%
}.
Near $T_c$ the deconfined plasma is thus the softer phase.
In a supercooled transition it lies ahead of the wall, giving $c_+<c_-$ and no folds.
Only a superheated transition, from the confined to the deconfined phase, places the softer phase behind the wall, with $c_+/c_-\lesssim\sqrt{0.08/0.01}\simeq2.8$ if the plasma behind the wall stays close to $T_c$.
This reaches at most the onset of the folds, as shown by the two green regions in \cref{fig:fold}.

Although the present analysis was developed for vanishing conserved density, the same hydrodynamic structure extends naturally to finite density, where a conserved charge lifts the competition.
For a barotropic CSS equation of state the charge does not enter the pressure, so profiles, wall matching, and folds are unchanged and the density simply follows the flow (see \myendmatterref).
Conservation of the charge current $J^\mu=nu^\mu$, implies that $n\gamma v$ is constant across the wall, where $n$ is the charge density.
Since $w=\mu n$ at $T=0$, with $\mu$ the chemical potential associated with the conserved charge, the two fluxes require exactly $\mu_+\gamma_+=\mu_-\gamma_-$.
For an ideal Fermi gas that becomes massive behind the wall, $c_+^2=1/3$ and
\begin{equation}
 c_-^2=\frac{k_F^2}{3(k_F^2+m^2)},\qquad
 \frac{\mu_-}{m}=\frac{c_+/c_-}{\sqrt{(c_+/c_-)^2-1}},
\end{equation}
where $k_F$ is the Fermi momentum. Hence, any hierarchy only requires $\mu_-$ to lie slightly above the mass threshold, within \SI{2}{\%} for $c_+/c_-\geq5$.
For instance, free nucleons at saturation density, $n_0\simeq\SI{0.16}{\femto\m^{-3}}$, have $m\simeq\SI{939}{\MeV}$ and $k_F\simeq\SI{263}{\MeV}$, so $m/k_F\simeq3.6$ and $c_-\simeq0.16$, the value of our benchmark.

Softness alone is however not sufficient.
For example, the nuclear liquid--gas transition connects two phases of nonrelativistic nucleons with comparable sound speeds and stays near the diagonal, as depicted in olive green in \cref{fig:fold}~\cite{Walecka:1974qa,Serot:1984ey}.
Realistic dense matter in neutron stars, on the other hand, can reach the hierarchies for which the fold develops.
Nuclear equations of state, e.g.\ in CompOSE~\cite{CompOSECoreTeam:2022ddl}, are soft near saturation and can become stiff, $c_s\simeq\numrange{0.7}{0.8}$, in neutron-star cores~\cite{Bedaque:2014sqa,Altiparmak:2022bke} (see teal region in \cref{fig:fold}) while quark matter approaches $c_s^2=1/3$ from below at high density.
Since the fold thresholds rise with $c_+$, stiff phases ahead of the wall lower the contrast required, with charged scalar condensates, $c_s^2\to1$, as an extreme example (see Refs.~\cite{Son:2000xc,Moore:2024gmt}).
A transition from stiff hadronic to quark matter thus develops folds if the quark phase is soft enough near the transition, $c_-\lesssim 0.35$ for $c_+=0.75$.

\paragraph{Outlook}

Our results expose a limitation of the standard description of bubble hydrodynamics: for sufficiently hierarchical sound speeds, $c_+>c_-$, the same wall velocity and transition strength can correspond to different fluid profiles, including flows of opposite orientation, a possibility hidden in the bag model.
The ambiguity reflects missing information and is resolved once the thermodynamic normalization of the two phases and the entropy production at the wall are fixed, which must be determined from the microscopic theory.

Whether this degenerate regime is reached is set by the equation of state.
At vanishing density the sound-speed hierarchy grows only logarithmically with the mass of the nonrelativistic species, and confinement yields $c_+>c_-$ only for superheated transitions.
Both hence reach at most the onset of the folds.
A conserved charge removes this obstruction, and stiff phases ahead of the wall lower the contrast required, making dense matter the natural arena for the non-universality identified here.
Extending the analysis to state-dependent sound speeds, conserved charges~\cite{Barni:2026cssfiniteDensity,Barni:2026FiniteDensity}, and wall dynamics beyond local thermal equilibrium will determine where it arises in realistic theories.

\begin{acknowledgments}
\paragraph{Acknowledgments}

G.B. is supported by the Deutsche Forschungsgemeinschaft under Germany’s Excellence Strategy - EXC 2121 Quantum universe - 390833306\@.
The research of E.M. at IPPP is supported by STFC\@.
E.M. and G.B. further acknowledge funding from the Spanish Research Agency (Agencia Estatal de Investigaci\'on, MCIN/AEI/10.13039/501100011033), the European Union
NextGenerationEU/PRTR, and ESF+ through the IFT Severo Ochoa Center of Excellence grant
CEX2020-001007-S, as well as grants
CNS2023-14453, CNS2023-145069, and PID2022-137127NB-I00\@.
\end{acknowledgments}

\paragraph{Data availability}

The custom code used for the hydrodynamic analysis and to generate the results presented in this work is publicly available at \href{https://github.com/hydrodex-pt/hydrodex}{\faGithub\ \texttt{Hydr\'oDex}}.

\bibliography{bibliography}


\clearpage
\appendix
\onecolumngrid
\makeatletter
  \par\frontmatter@title@above
  \phantomsection
  \label{end-matter}
  {\frontmatter@title@format End Matter\par}
  \vspace{11.5\p@}
\makeatother
\twocolumngrid

\section{Thermodynamic completion and LTE selection}
\label{app:thermodynamic-completion}

The hydrodynamic construction fixes the velocity and enthalpy profiles but does not determine the relative thermal normalization of the two phases.
For the CSS equation of state this missing information can be parametrized by
\begin{equation}
    \Psi_N
    \equiv
    \frac{w_-(T_N)}{w_+(T_N)}
    =
    \frac{
        (1+c_-^2)a_-T_N^{q_-}
    }{
        (1+c_+^2)a_+T_N^{q_+}
    }.
    \label{eq:psiN-def}
\end{equation}
For a microscopic theory, $\Psi_N$ is fixed by its thermodynamics and encodes, in particular, the relative thermal normalization or effective degrees of freedom of the two phases~\cite{Ai:2023see,Sanchez-Garitaonandia:2023zqz}.
For a given hydrodynamic solution, $T_+/T_N = r_w^{1/q_+}$, while energy-flux conservation across the wall gives
\begin{equation}
    w_+\gamma_+^2v_+
    =
    w_-\gamma_-^2v_-,
    \qquad
    \gamma_\pm=(1-v_\pm^2)^{-1/2}.
    \label{eq:energy-flux-app}
\end{equation}

Define the ratio of outgoing to incoming entropy flux by $\mathcal R\equiv{s_-\gamma_-v_-}/{s_+\gamma_+v_+}$.
Since $s_\pm=w_\pm/T_\pm$, \cref{eq:energy-flux-app} implies
$\mathcal R = (\gamma_+ T_+)/(\gamma_- T_-)$.
Local thermal equilibrium corresponds to entropy conservation, $\mathcal R=1$, while the second law of thermodynamics requires $\mathcal R\geq1$~\cite{Ai:2021kak,Ai:2023see,Krajewski:2024zxg,Krajewski:2026kcm}.

Using $w_\pm\propto T^{q_\pm}$, the value of the thermodynamic normalization required by a particular hydrodynamic profile is
$\Psi_N =\mathcal R^{q_-}\Psi_{N,{\rm LTE}},$ with
\begin{align}
    \Psi_{N,{\rm LTE}}
    &=
    \frac{\gamma_+^2v_+}{\gamma_-^2v_-}
    \left(
        \frac{\gamma_-}{\gamma_+}
    \right)^{q_-}
    \left(
        \frac{T_+}{T_N}
    \right)^{q_+-q_-} .
    \label{eq:psi-min}
\end{align}
Thus $\Psi_{N,{\rm LTE}}$ is completely fixed by the hydrodynamic profile and is precisely the value selected in LTE.

The expression of $\Psi_N$ also makes clear why the second law alone does not select a unique branch.
It only requires $\Psi_N\geq\Psi_{N,{\rm LTE}}$, whereas specifying the amount of entropy production fixes $\mathcal R$ and therefore fixes $\Psi_N$ uniquely for each candidate.

Suppose now that a point $(\xi_w,\alpha_N)$ lies in one of the coexistence regions of \cref{fig:fold}.
Let the distinct hydrodynamic solutions be labeled by $i$.
Each candidate has a different local wall state, enthalpy ratio, and consequently a different LTE normalization $\Psi_{N,{\rm LTE}}^{(i)}$.
We quantify their closest separation by
\begin{equation}
    \mathcal D_{\rm LTE}(\xi_w,\alpha_N)
    \equiv
    \min_{i<j}
    \left|
        \ln
        \frac{
            \Psi_{N,{\rm LTE}}^{(i)}
        }{
            \Psi_{N,{\rm LTE}}^{(j)}
        }
    \right|.
    \label{eq:D-LTE-app}
\end{equation}
Two genuinely distinct hydrodynamic solutions could therefore represent LTE states of the same microscopic theory only if $\mathcal D_{\rm LTE}=0$.

For the benchmark $(c_+, c_-)=(0.9, 0.16)$ used in \cref{fig:fold}, we reconstruct all competing solutions throughout the coexistence regions. 
Their values of $\Psi_{N,{\rm LTE}}$ remain distinct in the interiors of these regions~\cite{Barni:2026hydrodex}.
Fixing the microscopic value of $\Psi_N$ therefore selects at most one competing LTE solution.

\section{Finite-density extension}
\label{app:finite-density}

At finite density, hydrodynamics must also conserve a current
\cite{Mateos:2026moc,Rezzolla:2013dea},
\begin{equation}
    J^\mu=n u^\mu,
    \qquad
    \partial_\mu J^\mu=0,
    \label{eq:charge-current}
\end{equation}
which adds a third matching condition at the wall,
\begin{equation}
    n_+\gamma_+v_+=n_-\gamma_-v_-.
    \label{eq:number-matching}
\end{equation}
For a general equation of state, $p$, $e$, and $n$ depend on both $T$ and $\mu$, the relevant sound speed is $c_s^2=\partial_e p|_{s/n}$, and energy--momentum and charge matching must be solved together~\cite{Rezzolla:2013dea,Barni:2026cssfiniteDensity,Barni:2026FiniteDensity}.

\paragraph{Barotropic limit}

The problem simplifies if the pressure depends only on the energy density in each phase, $p=p(e)$.
The stress tensor then does not depend on $n$, so the velocity and enthalpy profiles and the wall matching of \cref{eq:matching} are those of the neutral problem.
For constant $c_s$, this includes the CSS classification and its folded branches.
Charge conservation determines the change in density along the flow without affecting it. 

For a smooth spherical self-similar profile, \cref{eq:charge-current} gives
\begin{equation}
    (\xi-v)\frac{\dd n}{\dd\xi}
    =
    n
    \left[
        \frac{2v}{\xi}
        +
        \gamma^2(1-\xi v)
        \frac{\dd v}{\dd\xi}
    \right],
    \label{eq:n-selfsimilar-app}
\end{equation}
which has the same right-hand side as the energy equation, with $n$ in place of $w$.
Hence $\dd n/n=\dd e/w$ and, for constant $c_s$, where $\dd w=(1+c_s^2)\,\dd e$,
\begin{equation}
    \frac{n(\xi_2)}{n(\xi_1)}
    =
    \left[
        \frac{w(\xi_2)}{w(\xi_1)}
    \right]^{1/(1+c_s^2)}.
    \label{eq:n-w-css-app}
\end{equation}
At the wall and at shocks, where the entropy per particle can jump, \cref{eq:number-matching} instead fixes the density on the downstream side.
The charge thus adds no constraint on the flow.
Constraints arise only once a microscopic equation of state relates $n$ to $T$ and $\mu$ in each phase.
As in the neutral case, $\Psi_N$ then determines which of the hydrodynamically allowed, possibly folded, solutions a given theory thermodynamically realizes.

\paragraph{Degenerate Fermi gas}

The example of the main text is an ideal gas of $4 N_{\!f}$  fermionic degrees of freedom with mass $m$ at $T=0$, which fill all momentum states up to the Fermi momentum $k_F$~\cite{Landau:1980mil,Kapusta:2006pm,Rezzolla:2013dea},
\begin{equation}
    n=\frac{2 N_{\!f} k_F^3}{3\pi^2},
    \qquad
    \mu=\sqrt{k_F^2+m^2} .
    \label{eq:fermi-gas-app}
\end{equation}
At $T=0$, $\dd e=\mu\,\dd n$ and $\dd p=n\,\dd\mu$, so the gas is barotropic, with $w=\mu n$ and
\begin{equation}
    c_s^2=\frac{n}{\mu}\frac{\dd\mu}{\dd n}=\frac{k_F^2}{3(k_F^2+m^2)}.
    \label{eq:fermi-cs-app}
\end{equation}
For fermions massless ahead of the wall and with mass $m$ behind it, $c_+^2=1/3$ and 
\begin{equation}
    \frac{c_+}{c_-}=\sqrt{1+\frac{m^2}{k_F^2}}\ ,
\end{equation}
with $k_F$ evaluated behind the wall.
Since $c_-$ varies with $k_F$, the CSS description applies locally around the wall state.
Dividing the energy-flux condition, $w_+\gamma_+^2v_+=w_-\gamma_-^2v_-$, by \cref{eq:number-matching} and using $w=\mu n$ gives
\begin{equation}
    \mu_+\gamma_+=\mu_-\gamma_-,
    \label{eq:mu-matching}
\end{equation}
so the state behind the wall is fixed algebraically by the incoming one, without the exponential suppression of the thermal case.
A ratio $r_c = c_+/c_-$ is reached for $\mu_-/m=r_c/\sqrt{r_c^2-1}$ , just above the mass threshold.

\end{document}